\documentclass[useAMS,usenatbib]{mn2e}
\usepackage{graphicx}

\newcommand{\Msun}{\ensuremath{\,{\rm M}_\odot}}                        % Solar mass symbol
\newcommand{\Rsun}{\ensuremath{\,{\rm R}_\odot}}                        % Solar radius symbol
\newcommand{\Lsun}{\ensuremath{\,{\rm L}_\odot}}                        % Solar luminosity symbol
\newcommand{\Teff}{\ensuremath{T_{\rm eff}}}                            % Effective temperature symbol

\title[Light curve analysis and orbital period change of the extreme mass-ratio overcontact binary AW~CrB]{Light curve analysis and orbital period change of the extreme mass-ratio overcontact binary AW~CrB}
\author[Eric Broens]{Eric Broens$^{1}$\thanks{E-mail:
eric.broens@skynet.be}\\
$^{1}$Vereniging Voor Sterrenkunde, Belgium}
\begin{document}

\date{Accepted 2013 January 17. Received 2013 January 16; in original form 2012 April 17 }

\pagerange{\pageref{firstpage}--\pageref{lastpage}} \pubyear{2012}

\maketitle

\label{firstpage}

\begin{abstract}
Extreme mass-ratio contact binaries with a high degree of overcontact may be in the late evolutionary
stages of the contact phase. Detailed photometric analyses and orbital period studies of those systems
can provide invaluable information for the coalescence scenario of close binary systems, as recently observed in V1309~Sco.
In this paper the first light curve analysis and period study for the totally eclipsing contact binary AW~CrB
is presented. The $VR_{c}I_{c}$ CCD photometric light curves are analysed by means of the WD code.
The asymmetry of the light curves is modelled by a cool star spot on the primary component. It is shown
that AW~CrB has a high degree of overcontact $f$~=~$75~\%$ and an extreme mass-ratio of $q$~=~0.10, placing
it among the few contact binaries with the lowest known mass-ratios. The mean density of the primary 
component suggest that it has evolved away from the ZAMS.\\
Eighty-four times of minimum light are reported, derived from observations available in 
public archives, the literature, and the observations presented in this paper. The orbital period
study shows a continuously increasing period, at a rate of ${{\mathrm{d}}P/{\mathrm{d}}t}~=~3.58\times10^{-7}~d~yr^{-1}$,
suggesting a mass transfer from the secondary component to the primary one, implying a further decreasing
mass-ratio.  
\end{abstract}


\section{Introduction}
The evolution of W UMa-type binary stars is still not well understood. The traditional view for the origin
of these contact binaries is that they are formed from detached cool binaries with initial orbital periods of a few days.
By a combination of evolutionary expansion of both components and angular momentum loss arising from the
magnetic stellar wind, a contact binary is formed. Model calculations suggest that these binary systems ultimately
coalesce into single stars and may be progenitors of the poorly understood blue straggler and FK Com–-type 
stars \citep[e.g. ][]{ras95, ste06, ste11}. It was recently demonstrated that the
eruption of V1309 Sco was the result of a merger of the components of a cool contact binary \citep{tyl11}. 
Extreme mass-ratio contact binaries with a high degree of overcontact may be in the late evolutionary stages
of the contact phase. Detailed photometric analyses and orbital period studies of those systems can provide
invaluable information for the evolution and coalescence scenario of these binary systems. 
 
The variability of AW CrB was discovered in the course of the ROTSE-I experiment, and classified as a
$\delta$ Scuti-type star with a period of 0\fd180465d \citep{ake00}. The variable identification, period
determination, and type classification were conducted by automatic algorithms. \citet{jin03}
found, from visual inspections of the light curves of the ROTSE-I $\delta$ Scuti stars, that several of these
stars were misclassified. In order to re-classify some of these stars, they started a multiband photometric
CCD observing programme for 20 of the 91 stars classified as $\delta$ Scuti star in the initial ROTSE-I
variable star catalogue. From their single night observations they found AW CrB to be a W UMa-type star with an
amplitude of about 0.3 mag. By combining the ROTSE-I data with their observations, they derived a period
of 0\fd360920. 
Their light curve of AW CrB displays total eclipses making this star very suitable for a photometric study.
The mass-ratio of W UMa-type stars exhibiting total eclipses can be accurately determined from photometric observations \citep{ter05}.
The long total eclipses of AW CrB, lasting about 10 per cent of the orbital period, in combination with the small amplitude
suggest a low mass-ratio making this star particularly interesting. 

The position of AW CrB furthermore
coincides with the {\it ROSAT\/} X-ray source 1RXS J161520.0+354218 \citep{ges06} indicating chromospheric activity.
Since no light curve model has been published yet, new photometric $VR_{c}I_{c}$ CCD observations were obtained.
In this paper the photometric analysis of these observations, using the Wilson-Devinney (WD) code \citep{wil71,wil72,wil79,wil94}, are presented
together with a period study. For the orbital period study also the observations published by \citet{jin03} and photometric
data obtained by the ROTSE and WASP robotic telescopes are used.

\section{Observations}

AW CrB was observed on eight nights in 2011 from a private observatory located near the town of Mol, Belgium, at an
altitude of about 40 meter above sea-level. The images were acquired using a 20-cm Celestron Schmidt-Cassegrain telescope,
equipped with a SBIG ST-7XMEI CCD camera, and $VR_{c}I_{c}$ filters which are close to the standard Johnson-Cousins system. 
The use of a focal reducer yields a field of view of 16.8\arcmin~by~11.2\arcmin~and a plate scale of 1.32\arcsec/pixel.
The operating temperature of the CCD was kept constant at a temperature about 30 degrees below the ambient temperature. For
most of the nights the operating temperature was kept constant at -20\degr C. On the second night the CCD was
cooled to -25\degr C and on the last two nights to -15\degr C. The observations were made under good
to excellent sky conditions, i.e. no sudden drops in the instrumental magnitudes, e.g. due to cirrus clouds,
have been observed. However, on JD 2455672 the observations had to be stopped early due to deteriorating weather conditions. 

The integration time for each image was 60 $s$. By using the CCD camera's additional autoguider chip, the stars were kept on
approximately the same pixels during an observing session. The FWHM of the stellar images was typically around 4 pixels,
occasionally increasing to more than 5 pixels partly due to a focus drift caused by the decreasing ambient temperature.

The images were processed with dark removal and flat-field correction using the imred packages in IRAF\footnote{{\sc IRAF}
is distributed by the National Optical Astronomy Observatories, which are operated by the Association of Universities for Research in
Astronomy, Inc., under cooperative agreement with the National Science Foundation.}.
Twilight flat-field images were obtained on all but two nights on which the flat-fields were obtained from a diffuse reflector
indirectly illuminated by an incandescent lamp. Vignetting is evident in the flat-field images with about 5 to 10 per cent loss
of intensity at the very corners of the image. The variable, comparison, and check star are well located around the central
area of the image which is not affected by the vignetting. 

Differential aperture photometry was performed using the digiphot/apphot package in IRAF. GSC 2586-1883 and GSC 2586-1807 were
chosen as the comparison and the check star respectively. The coordinates, $V$ magnitude, and colours of AW CrB, the comparison,
and the check star are listed in Table~\ref{tabCoo}. The magnitude and colours for the comparison and check star are taken from
the AAVSO Photometric All Sky Survey (APASS), which
is conducted in five filters: Johnson $B$ and $V$, plus Sloan $g'$, $r'$ and $i'$ \citep{hen12}. 
Using the equations from \citet{jes05}, the APASS Sloan magnitudes of the
comparison and check star have been transformed to $V-R_{c}$ and $R_{c}-I_{c}$ colours in order to transform the differential observations of
AW CrB to the standard Johnson-Cousins system. The comparison star was close enough to the variable that extinction differences
were negligible. The observations were made at airmasses smaller than 1.90.

\begin{figure}
\includegraphics[width=\columnwidth, angle=0]{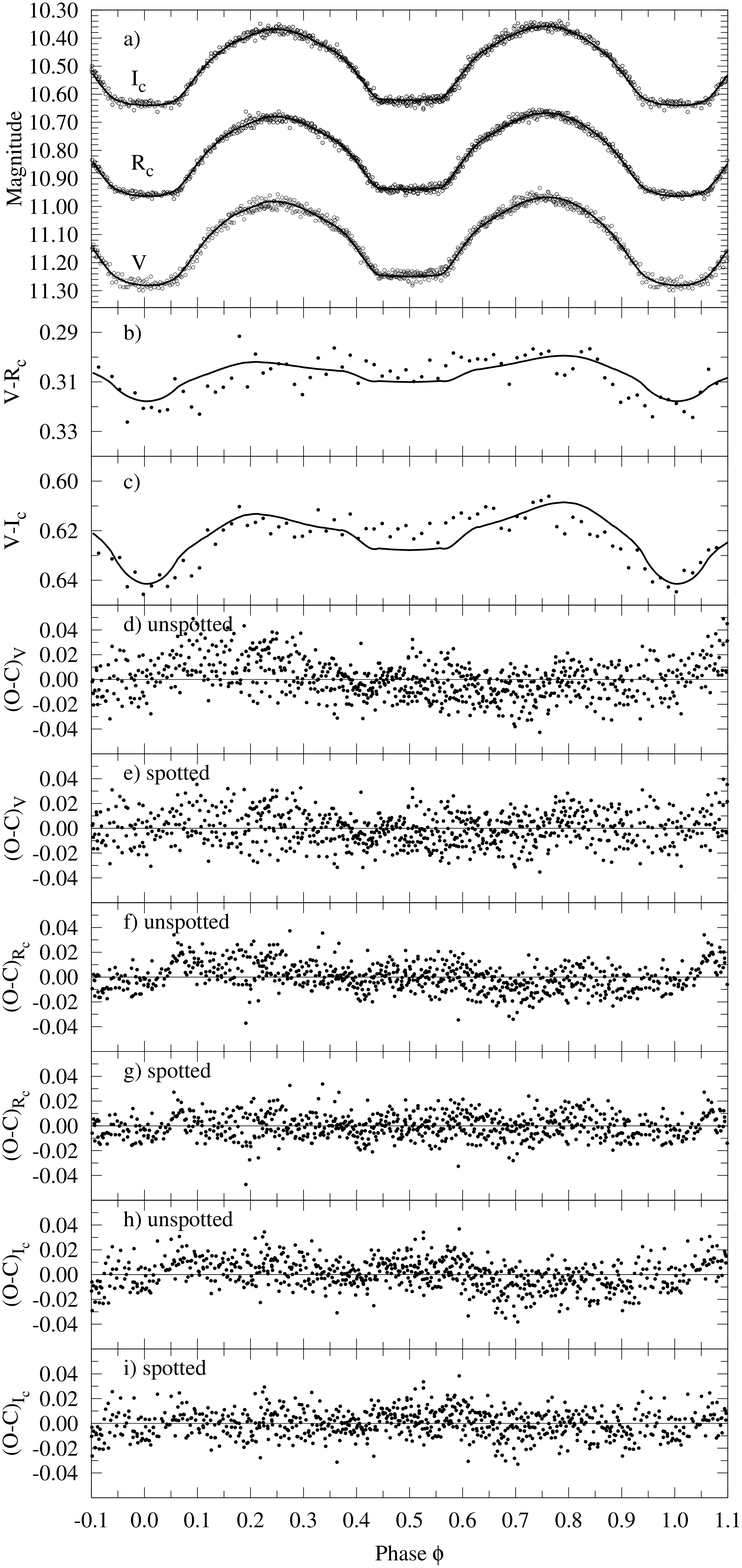}
\caption{
$VR_{c}I_{c}$ light curves (a) and colour curves (b and c) of AW CrB, the solid line is the synthetical light curve of the model with a cool spot on the primary.
The dots in the colour curves represent mean values calculated in a phase bins of $\phi~=~0.015$.
The lower panels display the residuals of the unspotted and spotted model fits for the $V$ (d and e), $R_{c}$ (f and g)
and $I_{c}$ passband (h and i) respectively.}
\label{lc}
\end{figure}

Depending on the airmass and the sky conditions, the integration time of 60 $s$ results in a signal-to-noise ratio for the $V$ filter  
between 251 and 521 for the variable, between 215 and 451 for the comparison star, and between 332 and 531 for the check star.
For the $R_{c}$ and $I_{c}$ filters the signal-to-noise ratio is somewhat higher. An estimate of
the uncertainty of the CCD photometry was obtained from the standard deviation of the differential light curve between comparison
and check star, viz. 0.01 magnitude in all filters. Since the photometric error is dominated by the fainter star, in this case
the comparison star, the above figure is a good estimate for the uncertainty of the variable minus comparison star light curve (e.g. \citet{how88}).
About 46~$h$ of CCD photometry was secured in each passband covering all orbital phases of AW CrB several times.
The complete observation log is given in Table~\ref{tabLog} and phased light curves are shown in Figure~\ref{lc}.
All observations are available from the AAVSO International Database\footnote{http://www.aavso.org} or upon request from the author.

\begin{table*}
\caption{The coordinates, $V$ magnitude and colours of AW CrB, comparison and check star.\label{tabCoo}}
\footnotesize
\begin{tabular}{lccccc}
\hline
Star           & $\alpha (2000.0)$                        & $\delta (2000.0)$              & $V$        & $V-R_{c}$ & $R_{c}-I_{c}$\\
\hline
AW CrB                     & $16^{\rmn{h}} 15^{\rmn{m}} 20\fs 2$     & $+35\degr 42\arcmin 26\farcs 2$ & $10.97$    & $0.30$    & $0.31$ \\
GSC 2586-1883 (comparison) & $16^{\rmn{h}} 15^{\rmn{m}} 21\fs 4$     & $+35\degr 50\arcmin 19\farcs 0$ & $11.73$    & $0.49$    & $0.46$ \\
GSC 2586-1807 (check)      & $16^{\rmn{h}} 14^{\rmn{m}} 56\fs 1$     & $+35\degr 50\arcmin 09\farcs 3$ & $10.43$    & $0.57$    & $0.53$ \\
\hline
\end{tabular}
\begin{flushleft}
For AW CrB the magnitude and colours at maximum light are listed, obtained from the observations presented in this paper. The magnitude and colours
for the comparison and check star are taken from the AAVSO Photometric All Sky Survey (APASS). 
\end{flushleft}
\end{table*}

\begin{table}
\caption{Observation log. \label{tabLog}}
\begin{tabular}{ccccc}
\hline
           &                  &       Number of data points        \\     
Julian day & Number of hours  &     $V$~~~$R_{c}$~~~$I_{c}$    \\
\hline
2455661    &   6.7            &     101~~101~~101      \\
2455664    &   6.3            &      92~~~91~~~91      \\
2455672    &   4.6            &      59~~~54~~~54      \\
2455676    &   6.9            &     104~~104~~104      \\
2455677    &   7.1            &     108~~107~~107      \\
2455685    &   3.9            &      64~~~65~~~65      \\
2455714    &   5.2            &      76~~~75~~~75      \\
2455715    &   4.8            &      72~~~70~~~70      \\
Total      &  45.5            &     676~~667~~667      \\
\hline
\end{tabular}
\end{table}

\section{Eclipse Timings and Orbital Period Study}

Up to now no times of minimum light had been published yet. From the observations reported in the present
paper two timings of primary minimum and five timings of secondary minimum were determined using the \citet{kwe56} method.
This method was also applied on the primary and secondary minimum observed by \citet{jin03} and
to the publicly available photometry from the WASP project \citep{but10}. To calculate the
time of minimum light from the data of the Northern Sky Variability Survey (NSVS) \citep{woz04},
the observations were first phased before applying the Kwee \& van Woerden method.
All timings of minimum light, spanning almost 12 years or about 12000 orbital revolutions, are listed in
Table~\ref{tabToM}. The timings from our observations and from \citet{jin03} are weighted means from the timings of minimum light
in the respectively 3 and 2 observation bandpasses.

\begin{table*}
\caption{Times of minima of AW~CrB. 
         \label{tabToM}}
\footnotesize
\begin{center}
\begin{tabular}{ccc|ccc|ccc}
\hline
Epoch    & HJD                     & data$^{\ast}$ & Epoch  & HJD                  & data$^{\ast}$ & Epoch  & HJD     & data$^{\ast}$\\
         & 2\,400\,000+            &     &        & 2\,400\,000+             &    &        & 2\,400\,000+            &     \\
\hline
-6730.5  & 51401.5390 $\pm$ 0.0050 & (1) & 1179.5 & 54256.5316 $\pm$ 0.0008 & (3) & 2058.0 & 54573.6121 $\pm$ 0.0007 & (3) \\
-6730.0  & 51401.7197 $\pm$ 0.0050 & (1) & 1182.0 & 54257.4317 $\pm$ 0.0016 & (3) & 2063.5 & 54575.5974 $\pm$ 0.0011 & (3) \\
-3908.5  & 52420.0945 $\pm$ 0.0003 & (2) & 1193.5 & 54261.5837 $\pm$ 0.0004 & (3) & 2072.0 & 54578.6654 $\pm$ 0.0007 & (3) \\
-3908.0  & 52420.2749 $\pm$ 0.0003 & (2) & 1196.0 & 54262.4863 $\pm$ 0.0005 & (3) & 2074.5 & 54579.5666 $\pm$ 0.0009 & (3) \\
   -0.5  & 53830.6248 $\pm$ 0.0013 & (3) & 1199.0 & 54263.5695 $\pm$ 0.0008 & (3) & 2077.5 & 54580.6502 $\pm$ 0.0009 & (3) \\
    5.0  & 53832.6097 $\pm$ 0.0003 & (3) & 1204.5 & 54265.5544 $\pm$ 0.0007 & (3) & 2088.5 & 54584.6208 $\pm$ 0.0010 & (3) \\
  146.0  & 53883.5033 $\pm$ 0.0021 & (3) & 1207.0 & 54266.4572 $\pm$ 0.0006 & (3) & 2094.0 & 54586.6078 $\pm$ 0.0006 & (3) \\
  997.0  & 54190.6585 $\pm$ 0.0005 & (3) & 1210.0 & 54267.5399 $\pm$ 0.0005 & (3) & 2127.0 & 54598.5157 $\pm$ 0.0011 & (3) \\
 1030.0  & 54202.5717 $\pm$ 0.0019 & (3) & 1212.5 & 54268.4422 $\pm$ 0.0008 & (3) & 2152.0 & 54607.5401 $\pm$ 0.0008 & (3) \\
 1033.0  & 54203.6534 $\pm$ 0.0008 & (3) & 1215.5 & 54269.5246 $\pm$ 0.0006 & (3) & 2157.5 & 54609.5275 $\pm$ 0.0022 & (3) \\
 1060.5  & 54213.5780 $\pm$ 0.0011 & (3) & 1218.0 & 54270.4269 $\pm$ 0.0005 & (3) & 2171.5 & 54614.5776 $\pm$ 0.0019 & (3) \\
 1066.0  & 54215.5646 $\pm$ 0.0008 & (3) & 1221.0 & 54271.5098 $\pm$ 0.0007 & (3) & 2188.0 & 54620.5367 $\pm$ 0.0010 & (3) \\
 1069.0  & 54216.6478 $\pm$ 0.0006 & (3) & 1226.5 & 54273.4943 $\pm$ 0.0011 & (3) & 2193.5 & 54622.5205 $\pm$ 0.0004 & (3) \\
 1074.5  & 54218.6327 $\pm$ 0.0012 & (3) & 1229.5 & 54274.5790 $\pm$ 0.0030 & (3) & 2218.5 & 54631.5396 $\pm$ 0.0011 & (3) \\
 1088.0  & 54223.5029 $\pm$ 0.0041 & (3) & 1232.0 & 54275.4787 $\pm$ 0.0010 & (3) & 2221.0 & 54632.4453 $\pm$ 0.0004 & (3) \\
 1091.0  & 54224.5872 $\pm$ 0.0019 & (3) & 1235.0 & 54276.5620 $\pm$ 0.0007 & (3) & 2235.0 & 54637.4993 $\pm$ 0.0009 & (3) \\
 1094.0  & 54225.6701 $\pm$ 0.0031 & (3) & 1240.5 & 54278.5481 $\pm$ 0.0012 & (3) & 2246.0 & 54641.4702 $\pm$ 0.0006 & (3) \\
 1099.5  & 54227.6553 $\pm$ 0.0014 & (3) & 1243.0 & 54279.4483 $\pm$ 0.0010 & (3) & 2271.0 & 54650.4924 $\pm$ 0.0010 & (3) \\
 1102.0  & 54228.5572 $\pm$ 0.0007 & (3) & 1248.5 & 54281.4339 $\pm$ 0.0011 & (3) & 2276.5 & 54652.4767 $\pm$ 0.0013 & (3) \\
 1107.5  & 54230.5426 $\pm$ 0.0007 & (3) & 1251.5 & 54282.5182 $\pm$ 0.0009 & (3) & 2287.5 & 54656.4483 $\pm$ 0.0007 & (3) \\
 1110.5  & 54231.6254 $\pm$ 0.0007 & (3) & 1257.0 & 54284.5026 $\pm$ 0.0004 & (3) & 2329.0 & 54671.4262 $\pm$ 0.0009 & (3) \\
 1116.0  & 54233.6118 $\pm$ 0.0010 & (3) & 1262.5 & 54286.4873 $\pm$ 0.0010 & (3) & 5072.0 & 55661.4760 $\pm$ 0.0003 & (4) \\
 1121.5  & 54235.5970 $\pm$ 0.0006 & (3) & 1268.0 & 54288.4732 $\pm$ 0.0004 & (3) & 5080.5 & 55664.5432 $\pm$ 0.0002 & (4) \\
 1124.0  & 54236.4998 $\pm$ 0.0009 & (3) & 1273.5 & 54290.4579 $\pm$ 0.0009 & (3) & 5102.5 & 55672.4850 $\pm$ 0.0006 & (4) \\
 1160.0  & 54249.4904 $\pm$ 0.0011 & (3) & 1279.0 & 54292.4442 $\pm$ 0.0005 & (3) & 5113.5 & 55676.4537 $\pm$ 0.0003 & (4) \\
 1163.0  & 54250.5731 $\pm$ 0.0007 & (3) & 1287.5 & 54295.5114 $\pm$ 0.0011 & (3) & 5116.5 & 55677.5359 $\pm$ 0.0003 & (4) \\
 1165.5  & 54251.4793 $\pm$ 0.0010 & (3) & 2016.5 & 54558.6344 $\pm$ 0.0008 & (3) & 5219.0 & 55714.5343 $\pm$ 0.0002 & (4) \\
 1168.5  & 54252.5618 $\pm$ 0.0013 & (3) & 2052.5 & 54571.6249 $\pm$ 0.0031 & (3) & 5221.5 & 55715.4363 $\pm$ 0.0002 & (4) \\
\hline
\end{tabular}
\begin{flushleft}
$^{\ast}$data: (1) NSVS \citep{woz04}, (2) \citet{jin03}, (3) WASP \citep{but10}, (4) this paper.\\
For minima observed in more than one passband the weighted mean of the timings in those passbands is listed.
\end{flushleft}
\end{center}
\end{table*}

\begin{figure}
\includegraphics[width=\columnwidth, angle=0]{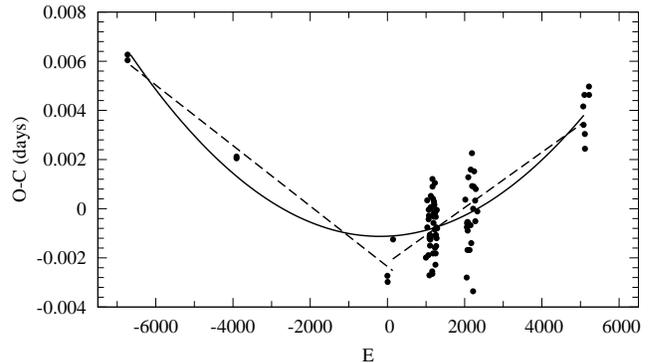}
\caption{
$O-C$ diagram of AW CrB. Dashed lines refer to a sudden period increase and solid line to a continuous period increase.}
\label{OC}
\end{figure} 

\begin{figure}
\includegraphics[width=\columnwidth, angle=0]{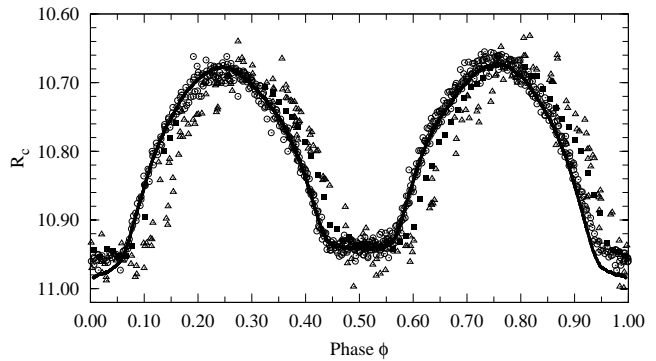}
\caption{Phase diagram of AW CrB constructed using the ephemeris in eq. (3). The solid line represents averaged
WASP observations. The circles, triangles, and filled squares are respectively the $R_{c}$ observations presented
in this paper, the NSVS observations, and the $I$ band data from \citet{jin03}.
The observations were shifted in magnitude to fit the observations presented in this paper.
\label{phase_diagram}}
\end{figure}

From the timings in Table~\ref{tabToM} the following linear least-squares ephemeris is obtained:
\begin{equation}
HJD~Min~I = 2453830.8080(3) + 0\fd3609353(1)\times{E}
\end{equation}
The $O-C$ residuals calculated with these ephemeris are plotted in Figure~\ref{OC}. 
The estimated errors of the minimum light timings, also calculated according \citet{kwe56}, cannot account for the scatter in the $O-C$ diagram.
These formal errors often underestimate the real error, due in part to random errors but there could also be biases arising from systematic differences
in the shape of the descending and ascending branches of the eclipse light curve.
Nevertheless, the $O-C$ diagram indicates an increase of the orbital period. The form of the period change is unclear due to the large gaps between
the groups of minimum light timings and the relative short time span of the observations. A sudden period increase and a constant period
increase are considered. 

If the period change is a sudden increase then with the data for $E \lid 146$ a least-squares linear fit yields
the ephemeris
\begin{equation}
HJD~Min~I = 2453830.8056(4) + 0\fd3609341(1)\times{E}
\end{equation}
With the data in the range of $-0.5 \lid E \lid 5221.5$ we obtain
\begin{equation}
HJD~Min~I = 2453830.8058(2) + 0\fd3609364(1)\times{E}
\end{equation}
With these equations a sudden period increase of $\Delta P~=~2.3\times10^{-6}~d$ or 0.2 $s$ is estimated to have occured
around $2005~-~2007$. Figure~\ref{phase_diagram} shows the phase diagram constructed with equation (3) for the WASP
data (solid line), the $R_{c}$ observations from this paper (circles), the \citet{jin03} $I$ band observations (filled squares),
and the NSVS data (triangles). The WASP, NSVS, and \citet{jin03} data were shifted in magnitude to fit the observations
presented in this paper. For the sake of clarity averaged WASP observations are plotted. 
The phase diagram demonstrates clearly the phase shift of the NSVS and \citet{jin03} data with respect to the WASP
data and the $R_{c}$ observations from this paper due to the period change.

Assuming a constantly increasing period, a quadratic least-squares fit to the light time minima yields
following ephemeris:
\begin{eqnarray}
HJD~Min~I = 2453830.8069(2) + 0\fd36093537(6)\times{E}\nonumber\\
+ 1.8(1)\times10^{-10}\times{E}^2
\end{eqnarray}
From the quadratic term of equation (4), a continuous period increase rate of
${{\mathrm{d}}P/{\mathrm{d}}t}~=~3.58\times10^{-7}~d~yr^{-1}$ or $0.03~s~yr^{-1}$ is derived.
However, due to the short time span of the observations, a cyclic period variation cannot be ruled out. 

The period analysis has been repeated using the timings of primary and secondary minimum separately. The results are
not significantly different from those obtained with both the primary and secondary minima included in the analysis.

\section{Light curve analysis}

\begin{figure}
\includegraphics[width=\columnwidth, angle=0]{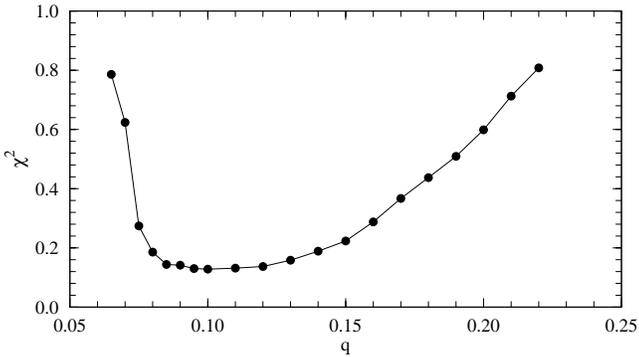}
\caption{Total $\chi^2$ for an unspotted model vs. a number of fixed mass-ratios $q$.\label{qsearch}}
\end{figure} 

The Wilson-Devinney (WD) method, as implemented in the software package {\sc phoebe} version 0.31a \citep{prs05}, was used to
analyse our $V$, $R_{c}$, and $I_{c}$ light curves simultaneously. All individual data points were used in the analysis.
Based on the colour indices $B-V = 0.505 \pm 0.024$, $V-R_{c} = 0.309 \pm 0.017$, $R_{c}-I_{c} = 0.286 \pm 0.005$, and interstellar reddening of $E(B-V) = 0.019$
published by \citet{ter12} and the 2MASS colours $J-H = 0.210$, $H-K_{s} = 0.055$, the temperature of star~1 (the star eclipsed at primary minimum, i.e. phase $\phi = 0$) was fixed at ${\Teff} = 6700~K$ \citep{cox00}.
Based on the temperature, convective atmospheres are assumed. The bolometric albedos $A_1 = A_2 = 0.5$
\citep{rus69} and the gravity-darkening coefficients $g_1 = g_2 = 0.32$ \citep{luc67}, appropriate for convective envelopes, were assigned. The logarithmic limb-darkening
law is used with coefficients adopted from \citet{vha93} for a solar composition star. The analysis is performed in mode 3, which is appropriate for 
overcontact systems that are not in thermal contact. A $q$-search method was applied to determine the initial mass-ratio. For a range of discrete values of $q$, the adjustable
parameters were: the orbital inclination $i$; the mean temperature of star~2, $T_2$; the monochromatic luminosities of star~1, $L_1$; and the dimensionless potential of
star~1 ($\Omega_1 = \Omega_2$, for overcontact binaries). From Figure~\ref{qsearch}, it can be seen that the resulting chi-square value of the convergent solutions reached its minimum for $q = 0.10$. 
At this point the mass-ratio $q$, with initial value 0.10, was included in the set of the adjustable parameters. The mass-ratio converged to a value of $q = 0.0992 \pm 0.0004$
in the final solution, with a total $\chi^2 = 0.129$. 
The solution reveals a high degree of overcontact $f = (\Omega_{in} - \Omega)/(\Omega_{in} - \Omega_{out}) = 75\% \pm 3$. 
The photometric parameters are listed in Table~\ref{tabPhotSol} and the residuals of the fit are plotted in Figure~\ref{lc} panels d, f, and h for the $V$, $R_{c}$,
and $I_{c}$ passband respectively. The errors given in this paper are the formal errors from the WD code and are known to be unrealistically small.

\begin{figure}
\includegraphics[width=\columnwidth, angle=0]{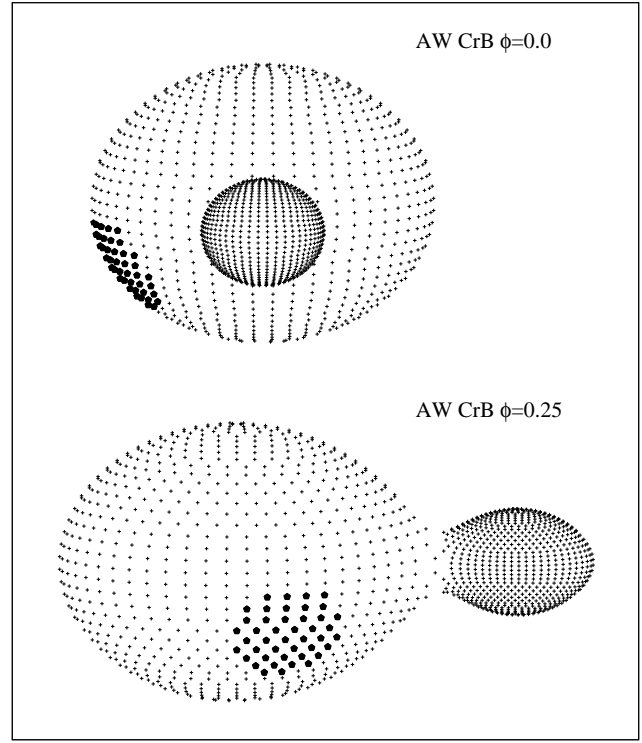}
\caption{
Geometrical representation of AW CrB at phases $\phi~=~0.0$ and $\phi~=~0.25$.}
\label{star_shape}
\end{figure}

While the overall fit of the computed light curves is quite satisfactory, Figure~\ref{lc} shows the maximum at phase $\phi = 0.25$ (Max I) slightly fainter than
at phase $\phi = 0.75$ (Max II) by about 0.02, 0.01, and 0.01 mag in the $V$, $R_{c}$, and $I_{c}$ bandpasses respectively. Table~\ref{tabMags} lists the average
magnitudes calculated in a phase interval of $\pm 0.02$ around the maxima and minima, with the standard deviation in units of the last digit given between parentheses.
The unequal light level between primary maximum and secondary maximum, the so-called
O'Connell effect, is known in many eclipsing binaries. It is often explained as surface inhomogeneities on either or both components and modelled by placing
circular dark or hot spots on the components.

We assumed that either a hot spot or a cool spot was on star~1 or star~2. For these four combinations and with the parameters from the spotless solution fixed, the parameter space
of colatitudes $\phi$, longitudes $\theta$, spot radius $r_{s}$, and temperature factor $T_{s}/T_{\star}$, with $T_{\star}$ being the local effective temperature of the surrounding photosphere, was explored using
the phoebe-scripter. The best solution was found for a cool spot on the primary, with $T_{s}/T_{\star} = 0.96$, spot radius $r_{s} = 21\degr$, colatitude
$\phi = 114\degr$, and longitude $\theta = 290\degr$. Since a solution
didn't converge with subsets of spot parameters made adjustable, these spot values were fixed while the higher mentioned set of star parameters was made adjustable
again. This resulted in a solution with a total $\chi^2~=~0.104$. All parameters are listed in Table~\ref{tabPhotSol} and the fit of the computed light and colour curves
is plotted in Figure~\ref{lc} panels a, b, and c.
The corresponding geometric structures at phases $\phi~=~0.00$ and $\phi~=~0.25$ are displayed in Figure~\ref{star_shape}. As shown by \citet{mac93}, spot determination by
photometry alone is unreliable because of the non-uniqueness of photometric solutions for spotted W~UMa-type binaries. The spot parameters listed in
Table~\ref{tabPhotSol} are therefore only provisional. 
Since \citet{pri08a} argue that perhaps all cool contact binaries are members of multiple systems, and a cyclic period variation cannot be ruled out, a third light
contribution to the light curves has been investigated. The analysis however did not reveal a third light contamination.
\citet{moc72} have shown that the internal contact angle for a given mass-ratio $q$, and inclination $i$, is almost independent of the degree of overcontact $f$.
The contact angle diagram plotted in their fig. 3 can be used as a consistency check on the solution found by the WD method. By estimating the internal contact
angle from the light curve, and by using the inclination found by the WD method, a mass-ratio $q~\sim~0.11$ is estimated from this diagram. This is in good agreement
with the solution found by the WD model. A significant difference would indicate third light or another complication.

\begin{table}
\caption{Average maximum and minimum magnitudes.\label{tabMags}}
\begin{tabular}{lcccc}
\hline
Filter              &  Max I             &  Max II            & Min I            & Min II          \\  
                    &  $\phi = 0.25$     &  $\phi = 0.75$     & $\phi = 0.0$     & $\phi = 0.50$   \\
\hline
$V$                 &  $10.99(1)$  &  $10.97(1)$  & $11.28(1)$ & $11.25(1)$ \\
$R_{c}$             &  $10.68(1)$  &  $10.67(1)$  & $10.96(1)$ & $10.94(1)$ \\
$I_{c}$             &  $10.37(1)$  &  $10.36(1)$  & $10.63(1)$ & $10.62(1)$ \\
\hline
\end{tabular}
\end{table}

\begin{table}
\caption{Light curve solutions of AW CrB.\label{tabPhotSol}}
\begin{tabular}{lcc}
\hline
Parameters              & unspotted    &  spotted       \\
\hline
$g_{1}=g_{2}$           & 0.32         &  0.32          \\
$A_{1}=A_{2}$           & 0.5          &  0.5           \\
$x_{1bol}$ = $x_{2bol}$ & 0.659        &  0.659         \\
$y_{1bol}$ = $y_{2bol}$ & 0.178        &  0.175         \\
$x_{1V}$ = $x_{2V}$     & 0.701        &  0.701         \\
$y_{1V}$ = $y_{2V}$     & 0.285        &  0.285         \\
$x_{1R}$ = $x_{2R}$     & 0.607        &  0.607         \\
$y_{1R}$ = $y_{2R}$     & 0.286        &  0.286         \\
$x_{1I}$ = $x_{2I}$     & 0.517        &  0.517         \\
$y_{1I}$ = $y_{2I}$     & 0.270        &  0.270         \\
$T_1$                   & 6700 K       &  6700 K        \\
$T_2$                   & 6769(11) K   &  6808(10) K    \\
$q$ = $m_{2}/m_{1}$       & 0.0992(4)    &  0.1012(4)   \\
$i$                     & 81\fdg6(3)   &  82\fdg1(1)    \\
$\Omega_1 = \Omega_2$   & 1.908(2)     &  1.913(2)      \\
$f$                     & 75\%(3)      &  75\%(3)       \\
$L_{1}/(L_{1}+L_{2})_V$ & 0.87(1)      &  0.87(1)       \\
$L_{1}/(L_{1}+L_{2})_R$ & 0.90(1)      &  0.90(1)       \\
$L_{1}/(L_{1}+L_{2})_I$ & 0.92(1)      &  0.92(1)       \\         
\\  
Spot parameters \\
$\phi$                  &              & 114\degr       \\
$\theta$                &              & 290\degr       \\
$r_{s}$                 &              & 21\degr        \\
$T_{s}/T_{\star}$       &              & 0.96           \\
\\
$\chi^2$                & 0.129       & 0.104         \\     
\hline
\end{tabular}
\end{table}

\section{Results and Conclusion}

The light curves obtained by CCD photometric observations in the $V$, $R_{c}$, and $I_{c}$ passbands were analysed simultaneously with the WD method, as implemented
in {\sc phoebe}. Since AW CrB shows total eclipses the photometric parameters can be determined reliably.
The photometric solutions suggest that AW CrB is an extreme low mass-ratio overcontact binary with $q = 0.10$ and a high degree of overcontact $f = 75\% \pm 3$. 
This places AW CrB among the few contact binaries with the lowest mass-ratios known up to now. The mass-ratio of the components, which is related to the angular
momentum loss and mass transfer, is one of the crucial parameters in the evolution of close binary systems. Table~\ref{LMRWUMa} lists the currently known
contact binaries with mass-ratios $q \la 0.12$.

\begin{table*}
\caption{Physical parameters for contact binaries with the lowest known mass-ratios.\label{LMRWUMa}}
\begin{center}
\begin{tabular}{lcccccccccccccc}
\hline
Star         & Type&   $M_{1}$ & $M_{2}$  &  $R_{1}$ &  $R_{2}$ & $L_{1}$  & $L_{2}$  & $T_{1}$  & $T_{2}$        & $q$         & $f$   & Ref.$^c$   \\
             &     &   (\Msun) & (\Msun)  & (\Rsun)  & (\Rsun)  & (\Lsun)  & (\Lsun)  &   (K)    &   (K)          &             & \%    &        \\
\hline
V857 Her$^a$ & A   &           &          &          &          &          &          &  8300    &    8513        &   0.0653    &   84  & (1)      \\
SX Crv       & A   &   1.246   &  0.098   &  1.347   &  0.409   &  2.590   &  0.213   &  6340    &    6160        &   0.0787    &   27  & (2)      \\
V870 Ara     & W   &   1.503   &  0.123   &  1.670   &  0.610   &  2.960   &  0.500   &  5860    &    6210        &   0.082     &   96  & (3)      \\
FP Boo       & A   &   1.614   &  0.154   &  2.310   &  0.774   &  11.193  &  0.920   &  6980    &    6456        &   0.096     &   38  & (4)      \\
DN Boo       & A   &   1.428   &  0.148   &  1.710   &  0.670   &  3.750   &  0.560   &  6095    &    6071        &   0.103     &   64  & (5)      \\
V1191 Cyg    & A*  &   1.29    &  0.13    &  1.31    &  0.52    &  2.71    &  0.46    &  6500    &    6610        &   0.105     &   74  & (6)      \\
CK Boo       & A   &   1.442   &  0.155   &  1.521   &  0.561   &  2.924   &  0.401   &  6150    &    6163        &   0.106     &   91  & (7)      \\
FG Hya       & A*  &   1.444   &  0.161   &  1.405   &  0.591   &  2.158   &  0.412   &  5900    &    6012        &   0.112     &   86  & (8)      \\
GR Vir       & A*  &   1.376   &  0.168   &  1.490   &  0.550   &  2.806   &  0.493   &  6300    &    6163        &   0.122     &   78  & (9)      \\
${\epsilon}$ CrA   & A   &   1.70    &  0.23    &  2.10    &  0.85    &  7.75    &  1.02    &  6678    &    6341        &   0.127     &   25  & (10)     \\
\\
AW UMa$^b$   & A   &   1.79    &  0.14    &  1.88    &  0.66    &  8.26    &  0.92    &  7175    &    7110        &   0.08      &   80  & (11)     \\
\hline
\end{tabular}
\end{center}
\begin{flushleft}
%\medskip
$^{\ast}$ Alternations between the A and W types have been reported for these stars. \\
$^a$ For V857 Her no spectroscopic mass-ratio could be determined yet. \citet{pri09} found a contribution from a early-type component in their
spectra and note that the shallownes of the eclipses might be caused by the 3rd light of this additional component and therefore put in doubt the low mass-ratio obtained
by light curve modelling.\\
$^b$ \citet{pri08b} found a spectroscopic mass-ratio of $q \simeq 0.10$ and found strong indications that the system is not a contact binary.\\
$^c$ References: (1) \citet{qia05b}; (2) \citet{szo04}; (3) \citet{sza07}; (4) \citet{gaz06}; (5) \citet{Sen08}; (6) \citet{ula12}; (7) \citet{kal05}; 
(8) \citet{qia05a}; (9) \citet{qia04}; (10) \citet{yan05}; (11) \citet{yan08}
\end{flushleft}
\end{table*}

\begin{table}
\caption{Periods and period change rates for low mass-ratio overcontact binaries.\label{tabPeriodChanges}}
\begin{tabular}{lcccc}\hline
Star             & Period    & ${{\mathrm{d}}P/{\mathrm{d}}t}$ & $P_{c}$    & Ref.$^c$  \\
                 & ($days$)  &  ($d~yr^{-1}$)                & ($yr$)       &       \\  
\hline
V857 Her         & 0.38223   &  $+2.90\times10^{-7}$         &              & (1)      \\
SX Crv           & 0.31660   &  $-1.05\times10^{-6}$         &              & (2)      \\
V870 Ara         & 0.39972   &                               &              & (3)      \\
FP Boo           & 0.64048   &                               &              & (4)      \\
DN Boo           & 0.44757   &                               &              & (5)      \\
V1191 Cyg        & 0.31338   &  $+1.3\times10^{-6}$          &              & (6)      \\
CK Boo           & 0.35515   &  $+3.54\times10^{-7}$         &   15.8$^{a}$ & (7)      \\
FG Hya           & 0.32783   &  $-1.96\times10^{-7}$         &   36.4$^{a}$ & (8)      \\
GR Vir           & 0.34698   &  $-4.21\times10^{-7}$         &   19.3$^{b}$ & (9)      \\
${\epsilon}$ CrA & 0.59143   &  $+4.67\times10^{-7}$         &              & (10)     \\
\\
AW UMa         & 0.43873   &  $-2.94\times10^{-8}$         &   17.6$^{b}$ & (11)     \\
\hline
\end{tabular}
\begin{flushleft}
%\medskip
$^{a}$ Cyclic period change attributed to magnetic activity cycles \\
$^{b}$ Periodic period change attributed to the presence of a third body \\
$^c$ References: same as in Table~\ref{LMRWUMa}
\end{flushleft}
\end{table}

The light curves displayed in Figure~\ref{lc} suggest that AW CrB is an A-type overcontact binary system according to the classification of \citet{bin70}.
However, we derived a secondary temperature of ${\Teff}~=~6808 \pm 10~K$, which is higher than that of the primary component.
A weak O'Connell effect is observed with Max I slightly fainter than Max II. The position of AW CrB
coincides furthermore with the {\it ROSAT\/} X-ray source 1RXS J161520.0+354218 \citep{ges06} indicating enhanced chromospheric and coronal activity. 
The O'Connell effect may be explained by a model with a cool starspot on the primary component. The best model is obtained with a cool spot on star~1,
with $T_{s}/T_{\star} = 0.96$, spot radius $r_{s} = 21\degr$, colatitude $\phi = 114\degr$, and longitude $\theta = 290\degr$.

\begin{figure}
\includegraphics[width=\columnwidth, angle=0]{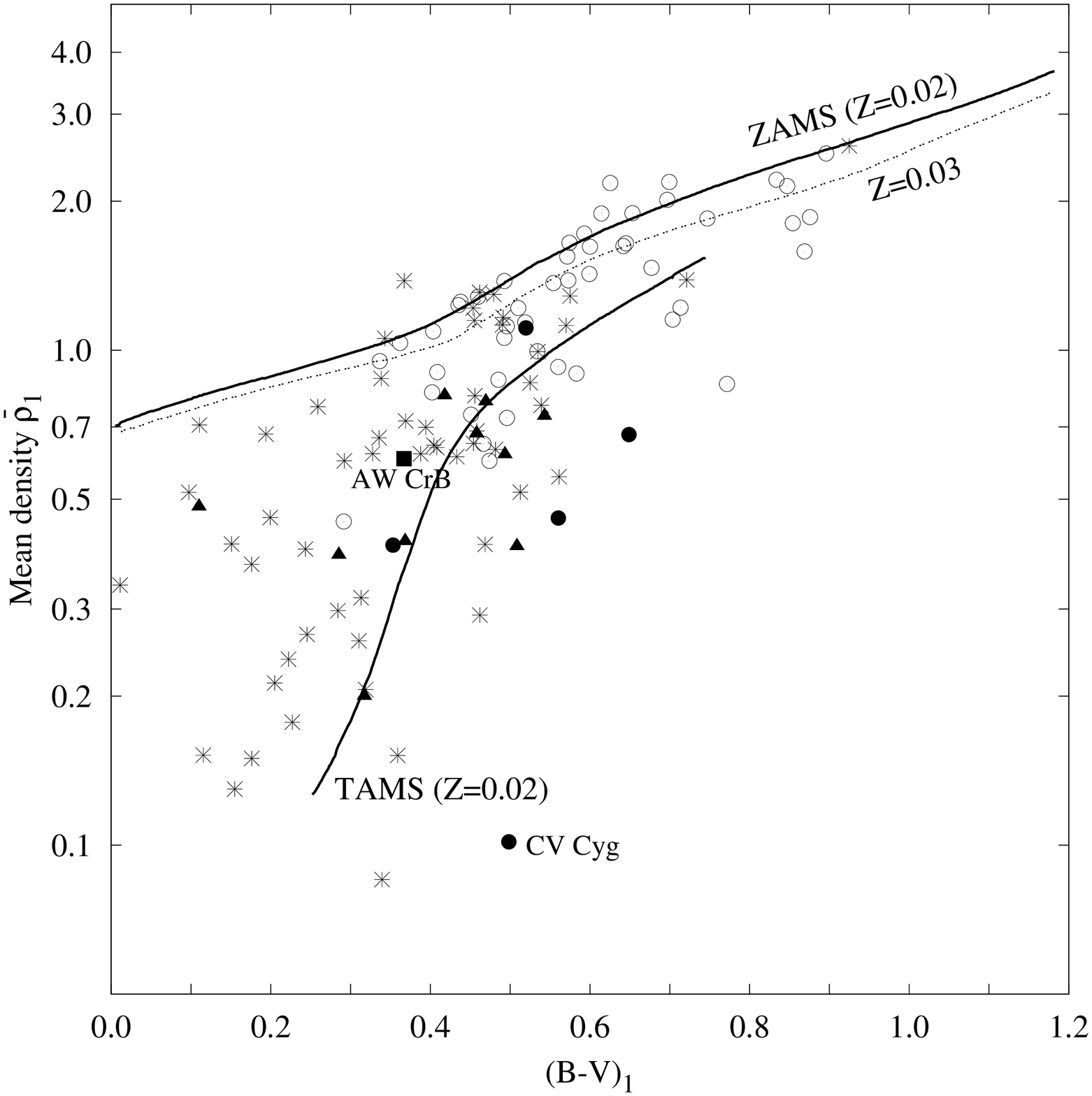}
\caption{
Corrected colour - mean density diagram for W-type (open circles) and A-type (crosses) primaries. The low mass-ratio W-type and A-type primaries are plotted with
filled circles and filled triangles respectively. AW CrB is indicated with a filled square. The zero-age main sequence and terminal-age sequence lines are taken from 
\citeauthor{moc81}'s \citeyearpar{moc81} Fig. 3.
}
\label{CDD}
\end{figure}

The evolutionary status of the primary component can be inferred from its mean density \citep{moc81,moc84,moc85}. Without knowledge of
the absolute dimensions the mean density $\overline{\rho_1}$, $\overline{\rho_2}$ of each component can be calculated with the formulae,
\begin{equation}
\overline{\rho_1}=\frac{0.079}{V_1(1+q)P^2}~g~cm^{-3},~\overline{\rho_2}=\frac{0.079q}{V_2(1+q)P^2}~g~cm^{-3},
\end{equation}
where $V_{1,2}$ are the volumes of the components using the separation $A$ as the unit of length, $q$ is the mass-ratio and $P$ the period in days. 
For AW CrB the mean densities $\overline{\rho_1}$, $\overline{\rho_2}$ can be determined to be 0.60 and 1.03, respectively. In the traditional contact binary models,
energy transfer from the primary to the secondary is assumed to explain the nearly equal temperatures of both components despite their considerable different
masses. Following \citet{moc81}, the colour of the primary can be corrected for this energy transfer in order to compare the density of the primary with
that of a zero-age main sequence star of the same spectral type. For AW CrB the corrected colour index ${(B-V)_1}=0.38$.
Figure~\ref{CDD} shows the position of the primary component of AW CrB in the ${(B-V)_1}$ - mean density diagram together with the primaries of other contact binaries.
The mean densities of the other primaries are calculated from data taken from \citet{pri03}. Contact binaries with components in poor thermal contact and hot
contact binaries, indicated respectively as type B and E in the catalogue, have been excluded. The zero-age main sequence and terminal-age
main sequence lines have been taken from \citeauthor{moc81}'s \citeyearpar{moc81} Fig. 3. The diagram indicates that the primary component of AW CrB has already moved
away from the ZAMS. This is the case for the majority of A-type contact binaries. The secondary's mean density $\overline{\rho_2}=1.03$ is nearly equal to the
density of a ZAMS star of the same spectral type. The evolutionary status of the secondary is however more difficult to judge as some of its properties, including the 
radius and thus the mean density, are possibly influenced by the energy transfer from the primary \citep{yan01}. 
In order to compare the evolutionary status of the primary components in A-type low mass-ratio contact binaries, the mean densities of the primaries for
the stars listed in table~\ref{LMRWUMa} are plotted in Figure~\ref{CDD} with a filled triangle. There is no indication that these primaries are
evolved in a greater or lesser extent than the primaries of A-type stars with a higher mass-ratio. For the W-type contact binaries, the primaries of the systems with
a mass-ratio $q \la 0.21$ have been plotted with filled circles. The diagram suggests that these are more evolved than the W-type members with a higher mass-ratio.
CV Cygni's primary has a very low mean density, which is also supported by its F8III spectral type. It should be noted however, that the \citeauthor{bin70}
classification of this star is uncertain \citep{vin96}.

The orbital period analysis based on observations collected from public available data of the NSVS and WASP project, the \citet{jin03} paper, and the present
paper, reveal an increasing period for AW CrB.
The form of this period change is unclear. Additional timings of minimum light over a longer time span are required to be conclusive. A sudden period increase of $\Delta P = 2.3\times10^{-6} days$
or $0.2~s$ may have occurred around $2005 - 2007$. However, continuous period variations are commonly encountered in contact binaries and are more
acceptable. Table~\ref{tabPeriodChanges} lists the period change rates for a number of extreme mass-ratio contact binaries.
In case of a continuously increasing period, the period change rate for AW CrB is estimated
to be ${{\mathrm{d}}P/{\mathrm{d}}t}~=~3.58\times10^{-7}~d~yr^{-1}$ or $0.03~s~yr^{-1}$.
Such a period change is usually attributed to mass transfer. If a mass of 1.4 {\Msun} is assumed for the primary component based on the colour indices, then a
conservative mass transfer rate from the less massive to the more massive component is estimated to be $5.1\times10^{-8} {\Msun}~yr^{-1}$.
With mass transferring, the orbital angular momentum decreases while the spin angular momentum increases. When the spin angular momentum of the system is more than a third
of its orbital angular momentum, this kind of low mass-ratio binaries with a high degree of overcontact may evolve into a rapidly rotating single star \citep{hut80}.

Due to the short time span of the observations, about 12 year or only 12000 orbital revolutions, a cyclic period variation cannot be ruled out. Periodic
variations, sometimes superimposed on a secular period change, have been reported for several W UMa-type stars. These periodic variations are usually explained by a
light-time effect via the presence of a third body, e.g. GR Vir \citep{qia04} and XY Leo \citep{yak03} among other stars. \citet{pri08a} argue that perhaps
all cool contact binaries are members of multiple systems, supporting the theory that Kozai cycles accompanied by the tidal friction may play an important role in
close binary formation \citep[and references therein]{ste11}.
In case period variations are not strictly periodic, they are usually explained by magnetic activity cycles in both components, e.g. FG Hya \citep{qia05a}
and CK Boo \citep{kal05}. The W UMa stars are fast-rotating solar-type stars and are known to show enhanced chromospheric
and coronal activity. \citet{app92} and \citet{lan98} proposed that the orbital period changes in close binaries are a consequence of magnetic
activity in one or both of the component stars.

Since the period variation of AW CrB is very important to understand its evolutionary state, it needs further long-term photometric monitoring for accurate
epochs of light minimum. In order to obtain the absolute parameters of AW CrB, and to check the derived photometric mass-ratio, spectroscopic observations are required.

\section*{Acknowledgements}

I gratefully acknowledge the AAVSO and the Curry Foundation for providing the CCD camera on loan. I thank Patrick Wils for proofreading
the draft of this paper and the referee, Professor S. W. Mochnacki, for suggesting several improvements to the paper.
This study used data from the NSVS created
jointly by the Los Alamos National Laboratory and the University
of Michigan, and funded by the US Department of Energy, the
National Aeronautics and Space Administration (NASA) and the
National Science Foundation (NSF). This research also used data from the WASP public archive. The WASP consortium comprises of the University
of Cambridge, Keele University, University of Leicester, The Open University, The Queen's University Belfast, St. Andrews University and the Isaac Newton Group.
Funding for WASP comes from the consortium universities and from the UK's Science and Technology Facilities Council.
Additionally this study made use of NASA's Astrophysics Data System, and the SIMBAD and VizieR databases operated at the CDS, Strasbourg, France.

\label{lastpage}
\end{document}